\documentstyle[emulateapj]{article}
\begin{document}

\submitted{Submitted 1999 Jan 19; accepted 1999 Jul 12; to appear in the Astronomical Journal (1999 Nov issue).}

\title{The Environments of Supernovae in Post-Refurbishment {\sl Hubble Space Telescope\/} Images\footnote{Based on observations made with the NASA/ESA {\sl Hubble Space Telescope}, obtained from the data archive of the Space Telescope Science Institute, which is operated by the Association of Universities for Research in Astronomy, Inc., under NASA contract NAS 5-26555.}}

\author{Schuyler D.~Van Dyk\altaffilmark{2}} 
\authoremail{vandyk@ipac.caltech.edu}

\author{Chien Y.~Peng\altaffilmark{3}}
\authoremail{cyp@as.arizona.edu}

\author{Aaron J.~Barth\altaffilmark{4}}
\authoremail{barth@exile.berkeley.edu}

\and

\author{Alexei V.~Filippenko}
\affil{Department of Astronomy, University of California, Berkeley, CA  94720-3411}
\authoremail{alex@wormhole.berkeley.edu}

\altaffiltext{2}{Current address: IPAC/Caltech, Mailcode 100-22, Pasadena, CA 91125.}

\altaffiltext{3}{Current address: Steward Observatory, University of Arizona, Tucson, AZ 85721.}

\altaffiltext{4}{Current address: Harvard/Smithsonian Center for Astrophysics, 60 Garden St., Cambridge, MA 02138.}

\begin{abstract}
The locations of supernovae in the local stellar and gaseous environment in 
galaxies contain important clues to their progenitor stars.  Access to this 
information, however, has been hampered by the limited resolution achieved by 
ground-based observations.  High spatial resolution {\sl Hubble Space Telescope\/} 
({\sl HST}) images of galaxy fields in which supernovae had been observed can 
improve the situation considerably.  We have examined the immediate environments 
of a few dozen supernovae using archival post-refurbishment {\sl HST\/} images.  
Although our analysis is limited due to signal-to-noise ratio and filter bandpass
considerations, the images allow us for the first time to resolve individual stars
in, and to derive detailed color-magnitude diagrams for, several environments.  We 
are able to place more rigorous constraints on the masses of these supernovae.
A search was made for late-time emission from supernovae in the archival images, and 
for the progenitor stars in presupernova images of the host galaxies.  We have 
detected SN 1986J in NGC 891 and, possibly, SN 1981K in NGC 4258.  We have also 
identified the progenitor of the Type IIn SN 1997bs in NGC 3627.  By removing younger 
resolved stars in the environments of SNe Ia, we can measure the colors of the 
unresolved stellar background and attribute these colors generally to an older, 
redder population.  {\sl HST\/} images ``accidentally'' caught the Type Ia SN 1994D 
in NGC 4526 shortly after its outburst; we measure its brightness.  Finally,
we add to the statistical inferences that can be made from studying the association 
of SNe with recent star-forming regions.
\end{abstract}


\section{Introduction}

A primary goal of supernova research is an understanding of the progenitor stars and 
explosion mechanisms of the different types of supernovae (SNe).  Unfortunately, a SN 
leaves few traces of the star that underwent the catastrophic event.  Inferences 
concerning the nature of the progenitor can be made from the ejecta at late times 
(Fransson \& Chevalier 1989; Leibundgut et al.~1991) and by probing the circumstellar emission from the 
blast wave (e.g., Fransson et al.~1989).  Unambiguous information can be derived if the 
progenitors can be directly identified in pre-explosion images, but to date this has been 
possible for only three SNe: SN 1987A in the LMC (West et al.~1987; Walborn et al.~1987, 1989), 
and probably SN 1978K in NGC 1313 (Ryder et al.~1993) and SN 1993J in M81 (Aldering, Humphreys,
\& Richmond 1994; Cohen, Darling, \& Porter 1995).  (The progenitor of SN 1961V has also been identified, 
but this event 
was most likely an $\eta$ Carinae-like outburst, rather than a genuine SN; see
Goodrich et al.~1989 and Filippenko et al.~1995a.)  In the absence of direct information about the 
progenitor stars, scrutiny of the host galaxies and local environments of SNe
continues to yield valuable clues to their nature.

SNe occur in at least three, and possibly four or more, spectroscopically distinct 
varieties (e.g., Harkness \& Wheeler 1990; Filippenko 1997).  Type I SNe are defined by the absence 
of hydrogen in their optical spectra.  The classical SNe~Ia are characterized by the 
presence of a deep Si~II absorption trough near 6150~\AA\ in their early-time photospheric
spectra ($t \lesssim 1$ month past maximum brightness).  Members of the Ib and Ic subclasses,
in contrast, do not show this line. Moderately strong He~I lines distinguish SNe~Ib from 
SNe~Ic.  SNe~II all exhibit hydrogen in their optical spectra, but the strength and profile 
of the H$\alpha$ line vary widely among these objects (e.g., Filippenko 1991; Schlegel 1996).
At late times ($\sim 5$--10 months past maximum brightness), SNe~Ia show strong blends of 
Co and Fe emission lines.  SNe~Ib and Ic, on the other hand, are dominated by relatively 
unblended lines of intermediate-mass elements, such as O and Ca, with SNe~Ic exhibiting 
larger line widths than SNe~Ib (Filippenko et al.~1995b).  SNe~II are dominated by the strong H$\alpha$
emission line, but otherwise they spectroscopically resemble SNe~Ib/c at late times.

Photometrically, the light curves of SNe I are all broadly similar, although SNe Ib and 
Ic are somewhat fainter and redder than SNe Ia, whereas those of SNe II exhibit much 
dispersion (Filippenko 1997, Figure 3; Patat et al.~1994).  SNe II are subclassified into 
``plateau'' (SNe II-P) and ``linear'' (SNe II-L), based on the shape of their light curves 
(Barbon, Ciatti, \& Rosino 1979; Doggett \& Branch 1985).

Previous ground-based studies of SN host galaxies and environments have primarily
concentrated on statistical results.  Presumably caused by the core collapse of massive 
stars, SNe~II have been associated with a young stellar population (e.g., Van Dyk 1992).  
The properties of SNe Ib and Ic favor massive progenitors which lose their hydrogen envelopes
prior to explosion (e.g., Porter \& Filippenko 1987; Filippenko 1997), but their progenitors have 
not yet been unambiguously identified.  Wolf-Rayet stars have been proposed (see Branch, Nomoto, \&
Filippenko 1991 and references therein), but a closer association with giant H~II regions {\it might\/} be 
expected for consistency with this hypothesis (Panagia \& Laidler 1991; Van Dyk, Hamuy, \& Filippenko 
1996).  Other possibilities include the off-center explosion of a white dwarf (Branch \& Nomoto 1986)
or the explosion of He stars in binary systems (e.g., Uomoto 1986).  The current paradigm of
white dwarfs as progenitors of SNe~Ia rests partly on their location in the parent galaxies.
SNe~Ia are not {\it clearly\/} concentrated toward star-forming regions (Maza \& van den Bergh 1976; 
Van Dyk 1992; McMillan \& Ciardullo 1996).  On the other hand, a correlation between the global H$\alpha$
brightness of the galaxy and the SN~Ia rate has been established (Oemler \& Tinsley 1979; 
Kennicutt 1984; van den Bergh 1991).  Bartunov, Tsvetkov, \& Filimonova (1994) conclude 
that SNe~Ia occur in spiral arms with a frequency very similar to that of the massive SNe~II,
but this is controversial.

Investigation of the local stellar and gaseous environments of SNe can, in favorable cases, 
yield useful constraints on the ages and masses of progenitor stars and therefore resolve 
ambiguities and contradictions in the progenitor models.  However, most studies of this kind
have been hampered by the limited spatial resolution of ground-based observations (e.g., 
Thompson 1982; Richter \& Rosa 1984; Huang 1987; van den Bergh 1991; Panagia \& Laidler 1991; 
Van Dyk 1992; Van Dyk et al.~1996).  The superior angular resolution of the 
{\sl Hubble Space Telescope\/} 
({\sl HST}) offers the potential for greater understanding of SN environments.  An 
investigation using data from the {\sl HST\/} archive was begun, as reported by Barth et 
al.~(1996), with the following goals: 1) to study the stellar populations in the 
immediate environments of SNe in cases where individual stars or clusters are resolved, and 
to use this information to help constrain the age and mass of the progenitor star; 2) to 
search for progenitor stars in images taken prior to SN explosions; 3) to determine whether 
old SNe are still visible; 4) to augment the ground-based data on the statistical association
of the different SN types with star-forming regions; and 5) to measure magnitudes for SNe
observed ``accidentally'' while still bright, adding to existing ground-based light curves.
We can also determine how sources in the fields of SNe affect the late-time SN light 
curves (Boisseau \& Wheeler 1991), and the photometry can be corrected for any contamination.

As Barth et al.~(1996) point out, although {\sl HST\/} images offer the tremendous advantage 
of higher angular resolution over similar ground-based data, the {\sl HST\/} images were 
almost exclusively obtained for purposes other than the analysis of SN environments, making 
it difficult to assemble a statistically meaningful sample (although, as the present study 
demonstrates, this issue becomes less severe as larger numbers of host galaxies are imaged).  
Additionally, one of the major limitations to a study of this kind is the small field of view 
of the {\sl HST\/} cameras; we find that many potentially useful images of SN host galaxies
do not contain the sites of SNe at all.  Also, exposure times are often chosen for the
purpose of imaging galactic nuclear regions, and the underexposed SN sites are not 
particularly useful.  Finally, many galaxies are imaged through only one {\sl HST\/} filter.

The environments of several SNe in pre-refurbishment HST images were studied by Barth et 
al.~(1996), but the results of this work were limited by the aberrated {\sl HST\/} optics 
and by the small size of the pre-refurbishment archival database.
Van Dyk et al.~(1999) recently examined the environment of the Type II-L SN 1979C in 
NGC 4321 (M100) in post-refurbishment {\sl HST\/} archive images.  As a result, 
they were able to place more rigorous constraints on the mass of the SN progenitor, which 
may have had an initial mass $M \approx 17$--18 $M_{\sun}$.  From additional {\sl HST\/} 
multi-band imaging obtained 17 years after explosion, they also recovered and measured the 
brightness of SN 1979C.

In this paper we report on the analysis of post-refurbishment {\sl HST\/} archive images 
of some additional SN environments.  As compared to the Barth et al.~sample, the number of 
environments is now significantly larger, to the point where we can begin to establish 
meaningful statistical results for the various SN types based on these {\sl HST\/} data.
Additionally and most importantly, a number of galaxies hae been imaged deeply and through 
two or more filters, primarily for Cepheid-based distance measurements, and several SN sites
are in these images.  Therefore, for the first time, we can obtain photometry of individual 
stars and small clusters in SN environments, in order to constrain the ages and masses of 
the progenitor stars.

\section{Archival Data and Photometry}

\subsection{The Sample}

We obtained from the {\sl HST\/} data archive all non-proprietary post-refurbishment
images of the host galaxies of all known SNe available by 1996 July 9, except when the 
nominal position of the image indicated that the SN site under consideration was definitely
outside of the field of view.  Additionally, some datasets were not considered because the
SNe had very poorly known positions or unknown spectroscopic types.  We also excluded from
our sample all SN environments imaged as part of the {\sl HST\/} Supernova Intensive Study 
(SINS) program (PI: R.~P.~Kirshner).  Only WFPC2 images from the archive met our selection 
criterion, and the vast majority were obtained as part of Cycle 5 programs.  All of the 
images had been processed by the routine WFPC2 calibration pipeline at the Space Telescope 
Science Institute (STScI).  

The images used in this study were taken on or after 1994 April 23, when the WFPC2 CCD 
temperature was adjusted.  Because most of the photometry we performed was in high 
background regions, the correction for any charge transfer efficiency effects would be on 
the order of, at most, a few percent, and we ignore these effects on the data.  Also, the 
near-IR quantum efficiency was stable for all observations (see Holtzman et al.~1995).  

Table 1 lists the SNe, their spectral types, host galaxies, WFPC2 detector containing the SN 
site, filters, observation dates, and total exposure times for the images which we have 
analyzed.  The spectral types were generally obtained from the online Asiago Observatory 
SN catalog (http://athena.pd.astro.it/$\sim$supern/snean.txt; see Barbon, Cappellaro,
\& Turatto 1989), although 
other sources, such as IAU Circulars and the Sternberg Astronomical Institute online catalog 
(http://www.sai.msu.su/sn/sncat/; see Tsvetkov \& Bartunov 1993), were also consulted.

\subsection{General Photometric Methods}

As can be seen from Table 1, a SN site is often found on one image taken through only one
filter, in which case we can only describe the general appearance of the SN environment.  
A single image, in most cases, actually consisted of two or more exposures, which were combined 
to remove cosmic rays, using the COMBINE task in the STSDAS package of IRAF\footnote{IRAF 
(Image Reduction and Analysis Facility) is distributed by the National Optical Astronomy 
Observatories, which are operated by the Association of Universities for Research in Astronomy, 
Inc., under cooperative agreement with the National Science Foundation.}.  In cases where the 
image consists of only a single exposure, we used a routine by M.~Dickinson which interpolates
around cosmic-ray hits to remove them from the image.  In other cases, images in more than one 
filter are available.  Of these, several cases discussed below allow us to resolve and perform 
photometry on individual stars and stellar objects in the SN environment.

When stellar objects could be measured in the SN environments, we employed point spread 
function (PSF) fitting photometry performed by DAOPHOT (Stetson 1987) and ALLSTAR within 
IRAF, which is designed to handle conditions of severe crowding, ill-defined local 
background, and low signal-to-noise (S/N) ratio.  Stars were generally located on the images 
using DAOFIND, with a detection threshold of 3$\sigma$, determined by the gain and read noise 
parameters for the image.

The principal element of the PSF fitting is an accurate model PSF, ideally derived from the 
same image as the SN environment, using a sufficient number of bright isolated stars.
This minimizes the problem of changing focus, chip distortions, spacecraft jitter, or 
pointing drifts, that causes PSFs to vary with time and position.  However, because of the
severe crowding and lack of stars of sufficient S/N ratio on the archival images 
we analyzed, it was generally impossible to build a good model PSF from field stars in these 
images.  Instead, we used the {\sl Tiny Tim\/} routine (Krist 1995) to produce an artificial 
PSF, simulated to have stellar spectral type A and PSF radius $0{\farcs}5$.  The artificial PSF
was created at the center of the region of interest, to minimize possible systematic
effects caused by the dependence of PSF shape on position.  A single model is sufficiently 
constant over the generally small ($\sim 10\arcsec \times 10\arcsec$) region to fit all 
stars.  In order to better match the width of our model PSFs with stars in the image, we 
convolved the Tiny Tim PSF with a Gaussian.  However, a subtle PSF mismatch would not contribute 
as substantially to photometric uncertainties as would low S/N ratio, crowding, and background 
brightness fluctuations.

In order to determine the possible systematic uncertainties in crowded field photometry, we added 
artificial stars with known brightnesses similar to our real objects.  Then, comparing their known 
magnitude with the outcome of the PSF fit we correct the real object magnitudes accordingly.  In 
general about $-$0.1 mag is added to correct for systematic uncertainties.

Count rates were converted to magnitudes using either the synthetic photometric zeropoints 
listed by Holtzman et al.~(1995; their Table 9) or Hill et al.~(1998; their Tables 4 and 5), 
appropriate for a $0{\farcs}5$ aperture radius.  (Much of the data we could analyze using 
PSF-fitting photometry were from the {\sl HST\/} Extragalactic Distance Scale Key Project, 
so for consistency we used the Hill et al.~zeropoints for these Key Project data.)  The 
Holtzman et al.~zeropoints were determined for the 14 $e^-$ ADU$^{-1}$ gain state, while the
Hill et al.~zeropoints were determined for the 7 $e^-$ ADU$^{-1}$ gain state.  A correction 
for the gain ratio to either set of zeropoints shows that they are quite consistent with 
each other.  When necessary, we corrected for throughput degradation, due to buildup of
contaminants on the camera optics.

The PSF-fitting photometry often resulted in color-color and color-magnitude diagrams (CMDs)
for the stars in the SN environment.  Throughout this paper we express the magnitudes
and colors in the WFPC2 synthetic magnitude system.  To analyze these diagrams, we used the
theoretical isochrones for solar metallicity from Bertelli et al.~(1994; hereafter B94), 
in order to constrain the ages and masses of the stars.  These isochrones were computed for
Johnson-Cousins $UBVRI$ and therefore require conversion into the WFPC2 synthetic magnitude 
system.  We produced conversions via synthetic photometry of model spectra, using the task 
SYNPHOT within IRAF/STSDAS and the Bruzual Synthetic Spectral Atlas within SYNPHOT.  In 
Table 2 we list our conversion formulae for the magnitudes and colors of interest (those 
listed in the tables in B94); the agreement with the Holtzman et al.~(1995) empirical 
conversions is generally quite good.  As noted in Holtzman et al., the passbands for the 
various filters in both filter systems are significantly different, and some filters, such 
as F336W, suffer from red leaks.  Magnitudes in the two filter systems tend to agree reasonably 
well for intermediate colors, but they diverge substantially for extremes in color.  Since 
many of the resolved stars are either very blue or very red, it is absolutely essential to 
perform these filter conversions.

Additionally, as noted in Holtzman et al.~(1995), because of the difference in passbands, the
extinction appropriate for the filters in each of the two systems differ.  We determined
the conversions for reddening and extinction in the Johnson-Cousins system to the WFPC2 
synthetic filter system, assuming the reddening laws of Cardelli, Clayton, \& Mathis (1989).  
These conversions are given in Table 3.

In some cases, where no resolved stars are apparent in the SN environment, we performed
aperture photometry, applying the magnitude zeropoint for infinite aperture.  The aperture 
photometry provides a larger-scale measurement of the magnitude, and where possible, color, 
of the general SN environment.  Additionally, in the case of some SN Ia environments where 
stars could be 	resolved and subtracted, we obtained the magnitude and color of the unresolved 
diffuse background through aperture photometry in order to place a lower limit on the ages 
of the old stellar populations.

Van Dyk (1992) and Van Dyk et al.~(1996) point out that a major factor affecting the results
of a study of this kind is the overall uncertainty in SN positions within their parent
galaxies.  This problem is particularly severe for these high-resolution {\sl HST\/} images.
In all cases, we verified the {\sl HST\/} image astrometry either against astrometry derived 
from ground-based images of the SN host galaxies or from careful visual inspection of available 
SN finding charts, or both.  We found that, especially when the {\sl HST\/} Fine Guidance Sensor 
lock mode was either ``coarse'' or ``gyro'', the image reference coordinates can be grossly in 
error (by 5 -- 10\arcsec\ or more).  In ``fine lock'' mode, in particular, image coordinates 
are generally good to, or better than, ${\sim}1{\farcs}5$--2\arcsec; for these images, the 
pixel positions of the SNe sites could be determined using the image header coordinate information, 
correcting for known WFPC2 geometrical distortion.  For the coarse-lock and gyro mode images, 
we estimated the pixel positions of the SN sites either by accurate offsets from a 
fiducial point, such as the galactic nucleus or an unsaturated star in the {\sl HST\/} image, 
or from available SN finding charts, or both.

In all cases, we assign an estimated error centered on the nominal SN position and regard the 
error circle as the SN environment.  This uncertainty radius varied from about $1{\farcs}5$ in 
the best cases, to 10\arcsec\ in cases where the SN positions are based only on nuclear offsets, 
adopting Marsden's (1982) estimate of these errors, even if the positional accuracy of the 
{\sl HST\/} image is good.  (It is our experience, from the work in Van Dyk et al.~1996, that the 
nuclear offset positions are actually not greatly discrepant with absolute positions; we believe 
that Marsden's estimate is a rather conservative, yet appropriate, one.)  It is a pity that, given
the high WFPC2 spatial resolution, SN sites cannot be more precisely defined in the {\sl HST\/} 
images.  But, even with this relative lack of precision, the fine image detail afforded by 
{\sl HST\/} in many of the SN environments considered here provides unprecedented information 
about these SN progenitors.

\section{The SN Environments}

Here we present and discuss the results for the individual SNe of each of the three
main SN types.

\subsection{The Type II SNe}

\subsubsection{SN 1926A in NGC 4303}

A pair of 80-s F606W WFPC2 exposures of NGC 4303 were available.  Three SNe II occurred in this 
galaxy: SN 1926A, SN 1961I, and SN 1964F.  Unfortunately, the site of SN 1961I is not on these 
images.  (We consider SN 1964F in \S 3.1.3.)  To find the site of SN 1926A on the 
{\sl HST\/} image, we determined the position of the nucleus of NGC 4303 and used the nuclear 
offset, 11\arcsec\ W and 69\arcsec\ N, given in the online Asiago SN catalog; the uncertainty
is assumed to be 10\arcsec.

Van Dyk et al.~(1996) found that this SN was associated with an H~II region of $\sim$4\arcsec\ 
radius, offset from the nucleus by 8\arcsec\ W and 70\arcsec\ N.
Within the large error circle, we find that the SN appears to have occurred near two spiral 
arms, with several small clusters and associations of stars near the site.  A small cluster
to the west of the SN's nominal position contains stars with F606W magnitudes which range
from about 23.4 to 24.4.  Assuming a distance modulus $m-M=30.91$ (Tully 1988), these 
imply $V \approx -7.5$ to $-$6.5, corresponding to bright (presumably massive) supergiant stars.

\subsubsection{SN 1940B in NGC 4725}

This galaxy was observed for the {\sl HST\/} Extragalactic Distance Scale Key Project, and the
archive contains a large number of F555W images, and a lesser number of F439W and F814W 
images.  Many more images in these bands were planned, but we have used only those available
before our cutoff date.  We combined the individual images into a single image in each band.
This galaxy has been host to two SNe: SN Ia 1969H, which is not in these images, and SN II 1940B,
whose environment is shown in Figure 1.  The position of this SN is not accurate (based on the 
nuclear offset, 95\arcsec\ E and 118\arcsec\ N), and we show the 10\arcsec\ error circle around
the nominal position.  The SN occurred near the edge of what appears to be two associations of 
bright, blue stars (also see Zwicky 1965, his Plate IV).  We performed PSF-fitting photometry 
on the combined images in all three bands, although we also analyzed the individual images, 
as a check on our results.

We show in Figure 2 the color-magnitude diagrams (CMDs) for the measured objects within
the error circle; most of them are in the bright associations.  Also shown are unreddened 
isochrones from B94 for various ages, corrected by the distance modulus $m-M=30.50$ to NGC 
4725 (Gibson et al.~1999).  The object at $m_{\rm F555W}\simeq 21.2$, F439W$-$F555W$\simeq$0.3, 
F555W$-$F814W$\simeq$0.6 mag, in particular, is too bright to be a single star and is probably a 
compact star cluster.  (Although we have left the isochrones unreddened, examination of the 
CMDs shows that reddening may be appropriate for some stars; SN 1940B itself may have been 
somewhat reddened; see Minkowski 1964, de Vaucouleurs 1974, and Patat et al.~1993.)

The environment of SN 1940B contains many bright, young, blue stars, particularly in the 
large OB associations.  Several red stars, presumably red supergiants, with possible ages 
of $\sim$8--20 Myr, are also seen on the CMDs.  The detected stars generally have ages,
depending on the actual reddening, in the range $\sim$6--30 Myr.  SN 1940B may have been 
associated with the populations of young stars; if the SN progenitor was a red supergiant, 
it may have had an age similar to the detected red supergiants, but could have been as young as 
6 Myr.  The lack of a precise SN position, in particular, prevents us from being more restrictive
about the nature of the progenitor.

\subsubsection{SN 1964F in NGC 4303}

We have already discussed in \S 3.1.1 this image, on which we also find the site of SN 1964F.  
This SN has been identified in the literature as a SN I, but based on R.~Barbon's re-examination 
of the SN spectra, the Asiago SN catalog lists it as a SN II.  From the nuclear offset in the 
Asiago catalog (28\arcsec\ W and 1\arcsec\ S), we derive a SN position, near which are several 
bright stars or clusters.  These stars have F606W magnitudes which range from about 23.1 to 
23.7.  Again, adopting $m-M=30.91$ (Tully 1988), these correspond to approximate $V$ 
magnitudes of $-$7.8 to $-$7.2, which are likely bright (massive) supergiants or, possibly, 
small clusters of stars (which may be true for the brightest objects).

\subsubsection{SN 1981E in NGC 5597}

Only a single F606W image of NGC 5597 is available.  
The SN position (uncertainty radius 10\arcsec) was derived from its nuclear offset (17\arcsec\ W
and 13\arcsec\ S).  The SN occurred near a spiral arm, with a number of bright stellar 
associations visible within the error circle, although the SN position does not closely coincide 
with any of them.

\subsubsection{SN 1981K in NGC 4258}

One pair each of a medium-band F547M and narrow-band F658N (H$\alpha$) and F502N ([O~III])
images of NGC 4258 are available.  SN was not spectroscopically classified, but is very
likely a Type II, based on the properties of its radio emission (Van Dyk et al.~1992).  It appears to 
have occurred along the edge of the northeast arm in this galaxy, near several large H~II complexes 
(Figure 3a).  Given the observed correlation of radio emission in SNe II with late-time optical 
emission (Chevalier \& Fransson 1994), we might expect SN 1981K to have still been a relatively luminous optical
emission source in 1995, when these images were made.  A very faint point-like source of possible 
H$\alpha$ emission is seen near the bottom of the error circle, with $m_{\rm F658N}=22.52{\pm}0.30$ mag
(adopting the STMAG zeropoint from Holtzman et al.~1995; their Table 9), which could be SN 1981K 
(or, possibly, a faint H~II region).  Using the online ``WFPC2 Exposure Time Calculator'' to convert
a point-source count rate into flux, we find $F_{{\rm H}\alpha}=2.0 \times 10^{-17}$ erg cm$^{-2}$ 
s$^{-1}$.  For a distance to NGC 4258 of 7.3 Mpc (Herrnstein et al.~1997), this corresponds to a luminosity of 
$L_{{\rm H}\alpha}=1.0 \times 10^{34}$ erg s$^{-1}$ (not corrected for reddening).  For comparison, 
Fesen et al.~(1999) recently estimated the dereddened late-time H$\alpha$ luminosity for SN 1979C 
to be $L_{{\rm H}\alpha}=1.5 \times 10^{38}$ erg s$^{-1}$.

If this is not the SN, but instead an unrelated emission source in the environment, then we can place 
an upper limit on the H$\alpha$ emission at the position of the SN:
$F_{{\rm H}\alpha}<2.5 \times 10^{-18}$ erg cm$^{-2}$ s$^{-1}$, or 
$L_{{\rm H}\alpha}<1.2 \times 10^{33}$ erg s$^{-1}$. 

We see from Figure 3b that the [O~III] emission in the field appears to be particularly associated  
with the cores of the bright H~II complexes.  
The [O~III] emission within the error circle, even at the position of the faint H$\alpha$ source,
is extremely faint.  If the H$\alpha$ emission is from SN 1981K, the relative lack of [O~III] emission
would be consistent with the shock/circumstellar matter interaction model by Chevalier \& Fransson 
(1994).

Figure 3c shows the continuum counterpart of the faint H$\alpha$ source, as well as some 
associated nebulosity or unresolved stars.  The source has $m_{\rm F547M}=22.36\pm0.08$ mag, or,
for the distance modulus $m-M=28.85$, $M_V \approx -6.5$.  A nearby small cluster of stars can 
also be seen inside the northwest edge of the error circle.  The brighter stars among them 
have $m_{\rm F547M}=23.07\pm0.18$, $22.84\pm0.08$, and $23.32\pm0.20$ mag; the faint star nearby has 
$m_{\rm F547M}=24.26\pm0.22$ mag.  These stars have $M_V \approx -5.8, -6.0, -5.5$, and $-4.6$ mag, 
respectively.  Although we do not have any color information for this environment, these magnitudes are 
consistent with those of supergiant stars.  The fact that little H$\alpha$ emission is associated
with these stars implies that they are non-ionizing, cooler, possibly yellow or red supergiants.  
Thus, SN 1981K may be associated with a population of evolved supergiants, with no recent 
star formation occurring in its immediate environment.  Given that the radio emission for the SN 
possibly arises from the SN shock interacting with the progenitor's red supergiant wind (Van Dyk et 
al.~1992), this would also imply that the SN's progenitor was a red supergiant of relatively low 
mass ($M \gtrsim 8 M_{\sun}$).

\subsubsection{SN 1982F in NGC 4490}

A pair of F606W images of the highly-inclined galaxy NGC 4490 were available.  From the nuclear offset 
(35\arcsec\ E and 20\arcsec\ S) for SN II 1982F (from the Asiago SN catalog), we derive a position for 
the SN near the edge of the PC chip.  SN 1982F occurred just outside the main body of NGC 4490.
The stars nearest the SN position have $m_{\rm F606W}=23.06\pm0.14$, $23.32\pm0.18$, and $22.92\pm0.07$ 
mag.  Given the distance modulus ($m-M=29.46$; Tully 1988), these correspond to $M_{\rm F606W}=-6.4$, 
$-6.1$, and $-6.5$ mag, and are probably bright supergiant stars.

\subsubsection{SN 1986J in NGC 891}

This SN II belongs to a peculiar class, known as SNe IIn (Schlegel 1990), which show narrow emission 
profiles atop broader bases in their spectra, particularly at H$\alpha$, possibly arising from the SN
shock interacting with very dense circumstellar matter.  SN 1986J was first discovered in the radio 
and subsequently followed optically (Rupen, van Gorkom, \& Gunn 1987; Leibundgut et al.~1991); 
it was missed in optical SN searches near the time of explosion.  The SN is a very luminous radio source 
(Weiler, Panagia, \& Sramek 1990) and should also still
be a luminous optical emission-line source.  Weiler et al.~(1990) provide an accurate radio position for
the SN, but, unfortunately, the coordinates for the pair of short-exposure F606W archival images of NGC 891
are greatly in error.  We located the SN site on the coadded image pair using a finding chart for SN 1986J 
(Rupen et al.~1987).  A star is seen in Figure 4 with $m_{\rm F606W}=21.28\pm0.06$ mag on 1994 Dec. 1, which
corresponds to the same object seen in the Rupen et al.~image, and is therefore almost certainly the SN.
Only faint, diffuse emission is visible around the SN, so it is not possible to study any associated 
stellar population.

\subsubsection{SN 1987K in NGC 4651}

Images in F218W, F547M, F555W and F814W are available for NGC 4651 in multiple exposures, except for the 
F547M band.  We used the finding chart for the SN IIb 1987K from Filippenko (1988) to aid in locating the 
SN environment on the images.  We show the SN environment in the F555W image in Figure 5.  Adopting an 
error of 2\arcsec\ in the SN position, we find that the SN occurred along a faint northern spiral arm in 
the galaxy.  No individual stars or clusters are detected within the error circle in the F555W and F814W 
images.  We measure a color F555W$-$F814W$\simeq 1.1$ mag for the SN environment, but we note that the 
environment appears dusty in both bands.  No emission in the SN environment is seen in the F218W image, 
but it is of a low S/N ratio.  In the F547M image a bright single star with $m_{\rm F547M}\simeq 24.2$
mag and some diffuse emission can be seen within the error circle.  For a distance modulus $m-M=31.13$ to 
NGC 4651 (Tully 1988), the star has $M_V \simeq -6.9$ mag.  This star could be the SN, seen at late times.
However, it is more likely a red supergiant star in the environment.  

\subsubsection{SN 1988M in NGC 4496B}

SN 1988M, discovered by Filippenko, Shields, \& Sargent (1988), occurred in NGC 4496B 
($cz=4510$ km s$^{-1}$), an optical double to NGC4496A ($cz=1730$ km s$^{-1}$), the host galaxy of SN Ia 
1960F (see \S 3.3.7).  NGC 4496A,B were imaged as part of a project to measure a Cepheid-based distance 
to SN 1960F (Saha et al.~1996a), so a large number of F555W and F814W images are available.  
These were combined to provide very deep images in both bands.  We show the SN environment in the F555W 
image in Figure 6.  The position is based on a nuclear offset ($12{\farcs}5$ S), measured accurately 
from a two-dimensional spectrum including the SN by Filippenko et al.~(1988); the uncertainty in the measurement 
is likely $\pm$3\arcsec.  However, comparing to ground-based images of these galaxies, the SN environment 
is very near the edge of the WFPC2 chip, so we conservatively assign an uncertainty of 5\arcsec\ to 
the SN position in the {\sl HST\/} image.  The SN appears to have occurred along a faint extension of a 
spiral arm in NGC 4496B.  Several very faint stars can be barely resolved by DAOPHOT in the environment, 
and they appear to be quite red; however, the extinction is unknown.  SN 1988M may have occurred among 
these red stars, either supergiants or reddened OB stars, with ages possibly as young as $\sim$20 Myr.  

\subsubsection{SN 1993G in NGC 3690}

NGC 3960 (Mk 171, Arp 299) has been host to perhaps five SNe in the past decade:  1990al (Huang et al.~1990), 
1992bu (Van Buren et al.~1994), 1993G, 1998T (see \S 3.2.5), and 1999D (see \S 3.1.12).  
SNe 1993G (Type II), 1998T (Type Ib/c), and 1999D (Type II) have been spectroscopically classified; 
the other two possible SNe (1990al and 1992bu) have been identified only in the radio and infrared, 
respectively.  A single short-exposure F606W image is available for NGC 3690.  Adopting the absolute 
position for SN 1993G from Forti (1993; we assign an uncertainty radius 2\arcsec), we find that the SN 
appears to have occurred well outside the main body of the interacting system, in a region of faint 
emission with no resolvable stars.

\subsubsection{SN 1997bs in NGC 3627}

A pair of archival F606W images of NGC 3627, which is also host to SN 1989B (see \S 3.3.13), were obtained.  
These images were taken well before the SN IIn 1997bs was discovered (Treffers et al.~1997).  Although 
SN 1997bs has an absolute position (Cavagna 1997), we used ground-based images, {\sl HST\/} images from 
later Cycles, and the discovery images to locate the SN environment, which we show in Figure 7a.  We are
able to assign a 1\arcsec\ radius uncertainty in the position.  Within the error circle is a star with 
$m_{\rm F606W}=22.86\pm 0.16$ mag on 1994 Dec. 28.  For a distance modulus $m-M=30.28$ (Saha et al.~1997), 
this corresponds to $M_{\rm V}\simeq -7.4$ mag for the star.  In Figures 7b and 7c we show the same field in
subsequent WFPC2 F555W images made as part of program GO 6549.  On 1997 Nov. 12 (Figure 7b), nearly seven 
months after the SN's discovery, the star's position is coincident with that of the SN, which had 
$m_{\rm F555W}=21.42\pm0.03$ mag.  On 1998 Jan. 10 (Figure 7c) the star had faded to 
$m_{\rm F555W}=23.37\pm0.05$ mag, below its original brightness.  Thus, we conclude that this star was the 
progenitor of SN 1997bs.  Although we have no color information for this star, its absolute magnitude is 
consistent with it having been an extremely luminous supergiant star.  This is only likely the fourth SN 
progenitor to be identified in pre-explosion images (see \S 1).  

Goodrich et al.~(1989) and Filippenko et al.~(1995) analyze the case of SN 1961V, concluding that it was 
a super-outburst of a luminous blue variable star, resembling the enormous eruptions sometimes experienced 
by $\eta$ Car. SNe with relatively similar spectra and low luminosities include SN 1999bw and SN 1997bs; 
they might be additional examples of such outbursts (Filippenko, Li, \& Modjaz 1999). Unlike normal SNe II, 
in which the progenitor destroys itself and creates a compact remnant (neutron star or black hole), here the 
progenitor survives the explosion reasonably unharmed.  Thus, in a sense these are not ``genuine'' SNe.  
The detection of a very luminous progenitor to SN 1997bs provides some evidence for this hypothesis, but the 
real test will be whether the star is still visible in future {\sl HST\/} images obtained years after the outburst.

\subsubsection{SN 1999D in NGC 3690}

The SN II 1999D recently occurred (Qiu, Qiao, \& Hu 1999) in the same interacting galaxy as SN 1993G
(\S 3.1.10).  The SN occurred near the outer edge of the galaxy, in a region of faint diffuse emission.
Within a 2\arcsec\ error circle around the SN's position are two unresolved star clusters, with 
$m_{\rm F606W}\simeq 20.2$ and 19.8 mag, respectively, along a more inward spiral arm.

\subsection{The Type Ib and Ic SNe}

\subsubsection{SN 1983V in NGC 1365}

Images of NGC 1365 in the F160BW, F336W, F547M, F555W, and F814W bands are available from 
two different programs, one of which is the {\sl HST\/} Extragalactic Distance Scale Key Project.  
Unfortunately, the Key Project images for this galaxy did not include the site of the SN Ic 1983V.  
The exposure times for the two sets of UV images were too short to be useful; we did not consider them 
further.  The F547M, F555W, and F814W images have reasonable S/N ratio after coaddition, such 
that we could produce limited color-magnitude diagrams for the SN environment.  The absolute position 
from Lindblad \& Grosbol (1983) was used to isolate the SN site, and we show the SN environment in Figure 8.
Van Dyk et al.~(1996) had found that the SN occurred within a bright H~II region of $\sim$5\arcsec\ radius.  
From the {\sl HST\/} image we see the stars and small clusters which are presumably ionizing the region.

We performed PSF-fitting photometry on the F555W and F814W images (the F547M band does not provide any 
additional color information and is less sensitive in the same wavelength range than the F555W band).  
In Figure 9 we show the resulting CMD.  The reddening to SN 1983V may be between $E(B-V) \simeq 0.18$ 
and 0.4 (Clocchiatti et al.~1997).  We assume both values of the reddening, applying them to the theoretical 
isochrones.  We also apply to the isochrones the distance modulus to NGC 1365 determined from Cepheids 
using {\sl HST\/}, $m-M=31.31$ (Silbermann et al.~1999).

The positions of the stars on the CMD in Figure 9 appear to agree better with the lower reddening than the 
higher.  One can see that the three objects in or very near the error circle around the SN position appear
too bright to be individual stars and are probably compact blue star clusters.  The resolved stars near 
the SN site have ages $\sim$4--10 Myr.  We can 
age-date the compact clusters by using the color evolution models for starbursts from Leitherer \& Heckman 
(1995).  In Figure 10 we show the colors of these probable star clusters, compared to the color evolution 
models for an instantaneous starburst from Leitherer \& Heckman, transformed to WFPC2 synthetic colors, 
and reddened by $E(B-V)=0.18$.  The line conventions in Figure 10 are the same as Figure 1 of 
Leitherer \& Heckman.

These clusters are evidently bluer than the model predictions.  Ideally, one would make this comparison 
having more color information.   However, we find that the cluster colors are consistent 
with very young starbursts, of ages $\lesssim$5 Myr.  If the progenitor of SN 1983V was associated with these 
clusters, then it may have had a comparably young age, implying a very high initial mass of $\sim 40\ M_{\sun}$.
Stars of this mass generally evolve to the Wolf-Rayet stage toward the end of their lives.  Thus, analysis of 
the environment of SN 1983V is consistent with the Wolf-Rayet progenitor model for SNe Ib/c, but the data are 
of limited sensitivity and color information.  

\subsubsection{SN 1985F in NGC 4618}

A pair of 80-s F606W exposures are available.  The image header coordinate information is 
erroneous, so we used ground-based images and the SN position given by Filippenko \& Sargent (1986) to isolate
the SN environment, which we show in Figure 11.  The SN occurred along the galaxy bar, on or near a bright 
H~II region (see Figure 1 in Filippenko et al.~1986).  We assign an uncertainty in the SN position of 2\arcsec\ radius.

Within this error circle are a number of bright stars.  We performed PSF-fitting photometry on the immediate 
field, including stars somewhat outside the error circle.  The brightest stars have $m_{\rm F606W} \simeq 20.4$ 
to 23.6 mag.  Assuming $m-M=29.31$ for the galaxy (Tully 1988), these correspond to $M_{\rm F606W} \simeq -9.0$ 
to $-5.8$ mag, which are consistent with, although somewhat brighter than, most luminous known supergiant and 
Wolf-Rayet stars.  The faintest stars detected within the error circle have $m_{\rm F606W} \simeq 25.2$ mag, 
corresponding to $M_{\rm F606W} \approx -4.2$ mag, consistent with the absolute magnitudes of red supergiants.

The brightest of these objects could be multiple stars in compact clusters, since the two brightest objects 
within the error circle have somewhat broader profiles than those of single stars in the image.  The brightest 
few objects could be multiple stars in compact clusters, since they have somewhat broader profiles than those of 
single stars in the image.  Unfortunately, we do not have any color information for this environment, to better 
constrain the stellar populations, and hence the age and mass of the progenitor.  The present evidence suggests 
that the progenitor of SN 1985F was a massive star, and is consistent with the exploding Wolf-Rayet star model 
(e.g., Begelman \& Sarazin 1986), but our data do not eliminate other possible models.

\subsubsection{SN 1990W in NGC 6221}

Only a single 500-s F606W image is available for NGC 6221.  We adopt the absolute position for the SN Ic 1990W
from the Sternberg SN catalog (Tsvetkov \& Bartunov 1993).  In Figure 12 we show the SN environment, with a 
2\arcsec\ radius 
uncertainty.  The SN occurred in a very bright cluster of stars to the northeast of the nuclear region.  We 
performed PSF-fitting photometry on these stars, taking care to distinguish cosmic-ray residuals from real stars.
The brightest star near the SN has $m_{\rm F606W}=19.29$ mag; the star closest to the SN position has 
$m_{\rm F606W}=20.94$ mag.  Assuming $m-M=31.44$ for the galaxy (Tully 1988), these correspond to 
$M_{\rm F606W} \simeq -12.2$ and $-10.5$ mag, respectively --- almost certainly too bright to be single stars.  
Fainter stars near the SN have $m_{\rm F606W} \simeq 23$--23.5 mag, which correspond to $M_{\rm F606W} \simeq -8.5$ 
to $-8.0$ mag, consistent with the brightest known supergiants and Wolf-Rayet stars.  
Unfortunately, we do not have any color information for this environment, but it appears that this SN occurred 
very near a massive, bright, and likely recent star formation region in NGC 6221, implying that the SN progenitor 
itself was also probably a young, massive star.

\subsubsection{SN 1996D in NGC 1614}

Only a single 500-s F606W image is available for NGC 1614.  We used the nuclear offset, $6{\farcs}6$ E 
(Drissen et al.~1996), with a 2\arcsec\ radius uncertainty, to isolate the environment of the SN Ic 1996D, which is 
shown in Figure 13.  The SN occurred along a bright spiral arm, in the vicinity of a number of stellar objects 
with median $m_{\rm F606W}\simeq 24.5$ mag and several unresolved objects with $m_{\rm F606W}\simeq 21.0$ to 20.8 
mag.  For a distance modulus to NGC 1614 of $m-M=34.02$ (assuming $cz=4778$ km s$^{-1}$ and $H_0=75$ km s$^{-1}$ 
Mpc$^{-1}$), these apparent magnitudes correspond to $M_V \simeq -9.5$ mag for the fainter objects, and 
$M_V \simeq -13.0$ to -13.2 mag for the brighter probable star clusters.  The image data do not allow us to be 
more specific about the stellar populations in the SN site, other than that it is likely these are recently-formed 
young, massive stars and star clusters.

\subsubsection{SN 1998T in NGC 3690}

The SN Ib 1998T occurred (Li, Li, \& Wan 1998) in the same interacting host galaxy as SNe 1993G and 1999D; we show the 
SN environment in Figure 14.  We tried to isolate the progenitor in this pre-SN image.  Unfortunately, given the 
distance to NGC 3690 ($\sim$42 Mpc, for $H_0=75$ km s$^{-1}$ Mpc$^{-1}$ and $cz=3121$ km s$^{-1}$) and the short 
exposure time, no individual stars can be resolved.  However, from the accurate position of the SN 
($11^h 28^m 33{\farcm}14$ $58\arcdeg 33\arcmin 44{\farcs}0$, J2000, from an $R$-band image taken with the Lick 
Observatory 0.75-m Katzman Automatic Imaging Telescope; see also Yamaoka et al.~1998) with a 2\arcsec\ radius 
uncertainty, we see that the SN occurred among the bright, dense concentrations of star formation occurring in 
this galaxy.  This single image of the SN environment implies that the SN progenitor was likely a massive star.

\subsection{The Type Ia SNe}

\subsubsection{SN 1895B in NGC 5253}

Several F255W images and a single 500-s F606W image are available for this galaxy  
before our cutoff date.  The galaxy has been the host of two bright SNe Ia, 1895B and 1972E.  
Unfortunately, the site of SN 1972E was not in either set of images.  We consider here the environment
of SN 1895B, adopting the absolute position (uncertainty 2\arcsec\ radius) of the SN from Caldwell 
\& Phillips (1989).  In the F255W image the SN is far from the very active star formation taking place
near the center of the galaxy, as traced by the UV-bright stars.  From PSF-fitting photometry of the 
F606W image (see Gorjian 1996) we find that the stars near the SN site have magnitudes 
$m_{\rm F606W} \simeq 24$ mag; the faintest among these have $m_{\rm F606W} \simeq 25.5$ mag.  The fact that 
these stars have no detectable counterparts on the deeper UV image implies that they are probably red.
Assuming the distance modulus ($m-M=28.08$) to NGC 5253 determined using Cepheids by Saha et al.~(1995),
these stars have $M_{\rm F606W} \simeq -4.1$ to $-2.6$ mag, consistent with bright red giants and fainter 
supergiants.

\subsubsection{SN 1937D in NGC 1003}

A pair of 80-s F606W exposures of NGC 1003 are available.  We used the nuclear offset position 
for SN 1937D from the online Asiago catalog (48\arcsec\ E and 1\arcsec\ S).  The SN appears to have occurred 
along a spiral arm between several stellar associations, although the inclination is high (68\arcdeg) for 
this galaxy.

\subsubsection{SN 1939B in NGC 4621}

Three images of NGC 4621 in each of F555W and F814W were obtained.  We used the nuclear 
offset position for SN 1939B (53\arcsec\ S) from the online Asiago catalog; it occurred toward the edge of 
the galaxy, away from the central region.  Several faint stars or clusters, as well as an unresolved 
background of stars, are within the large (10\arcsec\ radius) error circle.  Using PSF fitting we have 
subtracted the foreground stars and measured the color of the unresolved light within the error circle, which 
is F555W$-$F814W=1.37 mag, consistent with a background of K-giant stars.  It is therefore most likely that 
this SN~I was a SN Ia, rather than a SN Ib/c; the host galaxy is of early type, and the SN did not occur in 
or near a region of recent star formation.  The progenitor of SN 1939B, based on the color of the environment,
was likely old and evolved.

\subsubsection{SN 1940E in NGC 253}

A pair of F656N (Figure 15) and F675W images of NGC 253 were available.  We used the nuclear offset position 
for the SN I 1940E (51\arcsec\ W and 17\arcsec\ S) from the online Asiago catalog.  The SN occurred near the 
edge of dusty lanes or patches, among a field of diffuse starlight.  The fact that this SN site is not near a
dense stellar association or a region of H$\alpha$ emission makes it likely that the SN was a Type Ia.

\subsubsection{SN 1945A in NGC 5195}

A single 230-s F547M image and a pair of 700-s F218W images are available for NGC 5195.  
We used the nuclear offset position for the SN I 1945A (6\arcsec\ W and 4\arcsec\ S) from the online 
Asiago catalog.  Dust lanes cut across the diffuse starlight in the SN environment.  The deeper 
F218W image is completely blank in this field, indicating no detectable UV emission from stars or nebulae 
in this environment, and implying either a possible lack of recent star formation or appreciable extinction,
or both.  The extinction toward NGC 5195 has been estimated to be between $A_V=1.3$ (Thronson, Rubin, \& Ksir 1991)
and $A_V=2.0$ (Smith et al.~1990), at least partly due to dust from the arm of NGC 5194 obscuring the nucleus of NGC 
5195 (Zwicky 1957; Van Dyk 1987).  Based on the environmental evidence, it is not certain to which 
subclass SN 1945A belongs; however, its probable lack of association with massive star formation suggests 
a SN Ia.

\subsubsection{SN 1959E in NGC 4321 (M100)}

As is the case for SN 1979C (Van Dyk et al.~1999), the environment of the SN I 1959E in M100 can be analyzed
using the deep {\sl HST\/} Key Project images in F555W and F814W.  In Figure 16 we show the SN environment,
using the absolute position from Porter (1993).  The SN site is seen against a dust lane in this deep 
image.  Within the $1{\farcs}5$ radius error circle four stars are detected with magnitudes $m_{\rm F555W}$ 
from 24.1 to 25.5 ($-6.9$ to $-5.5$ for distance modulus $m-M=31.04$; Ferrarese et al.~1996) and colors F555W$-$F814W
from 0.1 to 1.3 mag (these stars are likely cool or reddened supergiants).  We performed PSF-fitting
photometry on the region of the galaxy beyond the error circle, out to $\sim$7\arcsec\ from the SN position;
in Figure 17 we show the resulting CMD.  We also show the B94 isochrones, reddened only by the Galactic 
foreground (Burstein \& Heiles 1984), although the reddening must clearly be larger than this.  A large number 
of populations are revealed from these deep images, from quite blue to very red, including hot main 
sequence stars, red supergiants, and some bright red giants.  Reddening and crowding are undoubtedly 
affecting the observed stellar magnitudes and colors.  However, we see that the photometry does not probe 
very deep in age (only to $\sim$100 Myr), in part because of the dust and the relatively bright
background.  Most of the measured stars appear to have ages $\sim$20--50 Myr.  After all resolved stars 
were subtracted away, we estimated the magnitude and measured the color of the diffuse background and find 
$m_{\rm F555W}\simeq 29.5$ mag (the magnitude of pixels 1$\sigma$ above the mean background) and
F555W$-$F814W$\simeq 0.7$ mag. This is consistent with a background of G stars, presumably giants with ages 
$\gtrsim 100$ Myr, or possibly fainter, reddened early-type stars (considering the extensive dust lanes in this 
environment).  From these results, we cannot rule out that SN 1959E was a SN Ib/c, with a massive
progenitor; however, we consider it more likely that it was a SN Ia, based on its environment.

\subsubsection{SN 1960F in NGC 4496A}

As is the case for SN 1988M in \S 3.1.9, the environment of the SN Ia 1960F is in the deep F555W and F814W
images, taken as part of a project to measure a Cepheid-based distance to SN 1960F (Saha et al.~1996a). 
Seventeen epochs of F555W imaging and four epochs of F814W imaging were obtained, each set of images 
consisting of a pair of 2000-s exposures.  The method for analyzing these images was the same as for 
the environment of SN 1979C (see Van Dyk et al.~1999): a model PSF was produced, stars were 
located, photometry was performed with DAOPHOT and ALLSTAR, and the results were averaged over all epochs.
Faint stars that were not found in the individual images were located in summed images, and the resulting
photometry was appended to the results for the individual images.  We performed the photometry over
a large area of the image.  Although crowding was severe in some regions, the photometry with a TinyTim
PSF led to satisfactory results.  The SN environment on the summed F555W image is shown in Figure 18.  
We used the accurate absolute position for the SN from Porter (1993).  The SN occurred in a region between
two spiral arms in NGC 4496A.

We show in Figure 19 the CMD produced by our photometry of the larger environment around the SN, along 
with the B94 isochrones, adjusted for both the reddening ($E[V-I]=0.04$) and the distance modulus 
($m-M=31.03$) from Saha et al.~(1996a), and transformed to WFPC2 synthetic magnitudes.  A large number 
of populations is revealed in these deep images, from quite blue to very red, including hot main 
sequence stars, red supergiants, and the tip of the red giant branch.  Reddening and crowding are 
undoubtedly affecting the observed stellar magnitudes and colors.  However, even these images are not deep
enough to see populations with ages $\gtrsim$500 Myr, from which SNe Ia are thought to arise.  After all 
resolved stars were subtracted away, we estimated a magnitude and measured a color for the environment, 
through a $1{\farcs}5$ aperture centered on the SN position, of $m_{\rm F555W}\simeq 30.1$ mag (the magnitude 
of pixels 1$\sigma$ above the mean background) and F555W$-$F814W$\simeq 0.6$ mag, consistent with a background 
of late F-stars, presumably giants with ages $\gtrsim 300$ Myr, or possibly fainter, reddened early-type 
stars.

\subsubsection{SN 1968A in NGC 1275}

Four F606W exposures of 3C 84 (NGC 1275) are available.  Using the nuclear offset, 7\arcsec\ E
and 24\arcsec\ S, from the online Asiago catalog, the site of the SN I 1968A is near the outer edge of 
the galaxy and the WFPC2 chip.  Only diffuse emission is seen near the SN site, and only two stellar-like
objects are seen within the 10\arcsec\ radius error circle.  These two objects have magnitudes 
$m_{\rm F606W}=20.2$ and 23.3 mag; for a distance of 70.2 Mpc (from $cz=5264$ km s$^{-1}$ and $H_0=75$ km 
s$^{-1}$ Mpc), these objects of $M_V\simeq -11.0$ and $-$14.0 mag are almost certainly compact 
clusters, more likely globular clusters.  We do not have any color information, but from the lack of 
detected recent star formation at the SN site, it is most likely that this SN was a Type Ia, rather than 
Type Ib/c.

\subsubsection{SN 1976D in NGC 5427}

Only a single F606W image of NGC 5427 is available.  We used the nuclear offset position for the
SN (35\arcsec\ E and 34\arcsec\ N) from the online Asiago catalog.  The SN site is very near the edge 
of the chip,  near a spiral arm.  Within the 10\arcsec\ radius error circle are a number of bright 
nebulae, presumably H~II regions and star clusters.

\subsubsection{SN 1981B in NGC 4536}

The site of this well-studied SN Ia is not in the deep Sandage et al.~{\sl HST\/} Cepheid images
(Saha et al.~1996b) of NGC 4536, but instead is in shallower F555W, F675W, F814W, and F658N exposures 
taken for another project.  SN 1981B occurred well beyond the galaxy's spiral arms, near several faint 
H~II regions (Figure 20).  We have performed PSF-fitting photometry of the stars in the three broad-band 
images within $\sim$10\arcsec\ of the SN site.  Unfortunately, the resulting CMDs do not go sufficiently 
deep to constrain the progenitor population for this SN Ia, since the oldest population detected is
apparently only $\sim 20$ Myr.  

After the stars were subtracted away from the broad-band images, we estimated the colors and magnitude of
the environment within the error circle.  We find that F555W$-$F675W$\simeq 1.6$ mag, 
F555W$-$F814W$\simeq 0.83$ mag, and $m_{\rm F555W}\simeq 28.45$ mag.  The F555W$-$F675W color almost 
certainly must be contaminated by H$\alpha$ emission in the environment.  The F555W$-$F814W color is 
consistent with late F-type stars, indicating a possible mixture of older red and younger blue stars in
the environment.

\subsubsection{SN 1983U in NGC 3227}

Only a single F606W PC image of NGC 3227 is available.  Using the absolute position of the SN Ia 1983U 
from the Sternberg SN catalog, we find that the SN site is along a faint extension of a spiral arm, far 
from the brighter stellar emission near the galaxy nucleus, and near (but not in) the dust lanes.

\subsubsection{SN 1988E in NGC 4772}

A pair of 80-s F606W exposures of NGC 4772 are available.  SN 1988E was possibly a peculiar SN I
(Pearce et al.~1990).  Using the absolute SN position from the Sternberg SN catalog, we find that 
the SN site is in a region of faint, diffuse emission on the outskirts of this elliptical galaxy, with no 
stellar objects within the 2\arcsec\ radius error circle.

\subsubsection{SN 1989B in NGC 3627}

A pair of F606W images of NGC 3627 were obtained (see \S 3.1.11).  We show the SN Ia 1989B environment 
in Figure 21, based on the absolute SN position from the Sternberg SN catalog.  The SN occurred along an 
inner spiral arm in an environment of bright stellar objects, diffuse emission, and dust, consistent with 
the relatively large reddening ($E[B-V]=0.37$ mag) found by Wells et al.~(1994) from the SN light curves.  
The resolved stars within the 2\arcsec\ radius error circle have $m_{\rm F606W}=22.0$ to 20.9 mag, which 
for a distance modulus $m-M=30.28$ (\S 3.1.11), correspond to $M_V\simeq -8.3$ to $-$9.4 mag; these are 
either extremely luminous single stars or, more likely, small compact star clusters.

\subsubsection{SN 1990N in NGC 4639}

Images in F218W, F300W, F439W, and F547M of NGC 4639 are available.  Unfortunately, the
deeper and more numerous Sandage et al.~Cepheid project images of this galaxy (Saha et al.~1997) do not contain 
the SN site, due to the site's proximity to a very bright foreground star.  However, from the images we 
obtained, using the absolute SN position from the Sternberg SN catalog, we find that the SN occurred near the
edge of the galaxy along an outer spiral arm.  The images in each band are not deep enough to resolve any
individual bright stars or clusters within the $1{\farcs}5$ radius error circle, nor for us to measure the 
colors of the environment.

\subsubsection{SN 1993R in NGC 7742}

Deep images in F336W, F555W, F675W, and F814W are available for NGC 7742.  SN 1993R is classified 
as peculiar Type I (Filippenko \& Matheson 1993), with substantial differences from the late-time (nebular) 
spectra of normal SNe Ia.  The spectrum of SN 1993R somewhat resembles that of the subluminous SN Ia 1991bg 
(Filippenko et al.~1992; Leibundgut et al.~1993), yet the Ca II near-infrared triplet is strong in SN 1993R 
and very weak in SN 1991bg.  The 
spectrum differs from the nebular spectra of SNe Ib and SNe Ic, in that the [O I] 6300 \AA\ emission is weak in 
SN 1993R.  In addition, the SN appeared to be photometrically subluminous near maximum brightness.

An approximate absolute position was given for SN 1993R by Treffers et al.~(1993).  We have measured a more 
accurate absolute position from an $I$-band image taken by B.~Leibundgut and W.~Vacca at the Lick Observatory 
Nickel 1-m telescope on 1993 August 18 UT.  Using the {\sl HST\/} {\it Guide Star Catalog\/} positions for 
foreground stars seen on the image, we derive a position of $\alpha=23^h 44^m 16{\farcm}29$, 
$\delta=10\arcdeg 46\arcmin 7{\farcs}1$ (2000.0).  The image header coordinate information for the {\sl HST\/}
images is erroneous, so, given the SN's absolute position and the absolute position (also from the Lick image) 
of an unsaturated foreground star on the {\sl HST\/} images, we located the SN site on the images
using the offset between the SN and the star.  We adopt a positional uncertainty of 1\arcsec\ for this offset.

We show the SN environment in Figure 22.  From the measured position, the SN occurred along the bright 
circumnuclear ring in the galaxy, in a region of patchy emission and dust.  Filippenko \& Matheson (1993) 
reported that the SN is superposed on a very bright H~II region.  This region appears as a
bright knot just outside the error circle for the SN in the {\sl HST\/} images and is not directly
coincident with the SN location.  This likely compact cluster has $m_{\rm F555W}=20.28$ mag, 
F336W$-$F555W=$-$1.29, F555W$-$F675W=0.71, and F555W$-$F814W=0.01 mag.  For a distance modulus $m-M=31.73$
(Tully 1988), it has $M_V\simeq -11.5$ mag and colors consistent with young, blue stars 
(the F555W$-$F675W color is almost certainly contaminated by H$\alpha$ emission).  Two stellar objects just 
outside the error circle have, respectively, $m_{\rm F555W}=22.52$ mag, F336W$-$F555W=$-$0.50, 
F555W$-$F675W=0.54, and F555W$-$F814W=0.71 mag; and, $m_{\rm F555W}=21.81$, F336W$-$F555W=$-$1.06, 
F555W$-$F675W=0.15, and F555W$-$F814W=0.11 mag.  The former object appears to be either a very bright 
red star or reddened compact cluster; the latter is likely a blue compact cluster.
Removing these stellar objects from the images and measuring the colors and magnitude for the environment,
we find $m_{\rm F555W}=18.25$, F336W$-$F555W=0.10, F555W$-$F675W=0.67, and F555W$-$F814W=0.93 mag.
These colors are significantly redder than the individual objects and are consistent with a mixture of
A through K stars, but also may be due to reddened fainter blue populations.

\subsubsection{SN 1994D in NGC 4526}

A set of short- and long-exposure F555W and F814W images are available.  SN Ia 1994D is 
clearly seen on the {\it HST\/} images; it occurred along the edge of the central dusty disk of this host 
galaxy, just beyond or north of the prominent dust lanes (Figure 23).  Although the SN is saturated in the 
long-exposure images, only the central pixels of the SN profile are saturated in the shorter exposures.  

We have measured the magnitudes of the SN in the two {\sl HST\/} filters by using the wings of the stellar 
profile and interpolating the flux lost in the saturated portion of the profile.  We used the unsaturated 
magnitudes of the field star (which is ``star C'' in Richmond et al.~1995), also seen on the image, as a 
reference.  We find $m_{\rm F555W}=14.48 \pm 0.08$ and $m_{\rm F814W}=14.19 \pm 0.08$ mag for SN 1994D on 
1994 May 9 UT.  These values agree well with $V=14.40$ and $I=14.19$ mag for 1994 May 8 UT by Richmond et 
al.  (Our measurements for the field star magnitudes, $m_{\rm F555W} = 16.35$ and $m_{\rm F814W} = 15.51$
mag, also agree well with Richmond et al.'s magnitudes for this star.)

The SN environment is dominated by diffuse starlight in the host galaxy.  We averaged the measurements of the 
magnitude and color of the SN environment on the deeper images through two $0{\farcs}5$-radius apertures, 
placed on either side of (and outside) the wings of the saturated SN PSF.  We find that $m_{\rm F555W}=19.40$ 
and F555W$-$F814W=1.52 mag in the environment, which is quite red.  Undoubtedly, dust must be contributing to 
the colors of this environment, but taken at face value, the color implies the presence of K-type stars, 
presumably giants, in the SN environment in this early-type host galaxy.

\section{Discussion and Conclusions}

We have conducted an analysis of SN environments in post-refurbishment {\sl HST\/} archival data.  Due to the 
superior resolution of {\sl HST}, the results of this study are a significant improvement over those involving 
ground-based data or our previous study of pre-refurbishment {\sl HST\/} data (Barth et al.~1996).  We have 
accomplished several of our main goals set out for this project.  (1) For the first time, through deep 
multicolor imaging undertaken by the {\sl HST\/} Cepheid projects, in particular, we have been able to produce 
color-color and CMDs for several SN environments, providing constraints on the ages and masses of stellar 
populations associated with the SNe, and allowing us to indirectly infer the ages and masses of the SN 
progenitors; this is true for SNe 1940B, 1983V, 1987K, and, to a lesser degree, 1988M.  (2) We have identified 
the stellar progenitor of SN 1997bs, clearly visible on a pre-SN {\sl HST\/} image, and we find that the 
progenitor was a luminous supergiant star.  (3) We have searched for old SNe on the images, being successful in
the cases of SNe 1979C (see Van Dyk et al.~1999), 1986J, and possibly 1981K, as expected, since these three objects 
showed evidence of interaction with circumstellar gas.  If the late-time light curves of other SNe declined 
exponentially, many or most should be well below detectability.  Also, several SNe had substantially more 
complicated environments and imprecise positions, thwarting our attempts to detect them at late times.  (4) We 
can, based on the archival data, advance the statistical association of the different SN types with star-forming 
regions.  (5) In the case of SN 1994D, which was caught ``by accident'' on {\sl HST\/} images, we can measure its
brightness, augmenting existing ground-based light curves for the SN.

The immediate environments of four SNe Ia considered above contained bright clusters and H~II regions with which 
these SNe Ia might have been associated.  However, Van Dyk (1992) has shown that SNe Ia are more likely chance 
superposed on recent star formation regions than physically associated.  We expect that SN Ia progenitors, based
on the current models, should be considerably older and fainter.  In three cases, we were able to do 
detailed analyses, based on CMDs, of the younger stars in the environments.  Removing the younger stars, through 
PSF fitting, allowed us to measure the colors of the underlying diffuse stellar emission, which may be due to 
populations similar to those of the SN Ia progenitors.  The colors of the diffuse emission provide suggestive 
indications that the stellar populations are generally old and red, consistent with red giants.  Since we are 
unlikely to measure the intrinsically faint individual stars in the stellar populations associated with SN Ia 
progenitors, regardless of the depth of the {\sl HST\/} exposures (except possibly for the cases of Local 
Group galaxies), we hope in the future to continue to amass statistics which can place more stringent constraints 
on the nature of SN Ia progenitors.

The four SNe Ib and Ic that we have included in this study seem more closely associated with brighter, more 
massive stellar regions than is the case for the SNe II in our sample, which tend to be associated with smaller, 
fainter stellar associations.  Van Dyk et al.~(1996), from their ground-based survey, found no difference in the
association of the two SN types with massive stellar regions.  We caution that the conclusions in the current 
paper are based on very few SNe Ib/c and a small number of SNe II; we clearly need to accumulate better 
statistics to confirm or refute this trend.  But, taken at face value, the high-resolution {\sl HST\/} images 
suggest that the SNe Ib/c progenitors may be more massive, in general, than the SNe II progenitors.  Exploding 
Wolf-Rayet stars have been proposed as possible progenitors for SNe Ib/c, and our limited results are consistent
with this scenario.

Fortunately, the environments of several SNe in our study have recently been imaged as part of other {\sl HST\/} 
projects, for longer exposure times and in various bands.  We will continue to acquire images for these, and 
other environments, as they become available in the future and will build on the current study in future papers.
With the ever-increasing number of images filling the {\sl HST\/} archive, we intend to enhance our sample 
significantly. Moreover, as new relatively nearby SNe occur, we will hunt directly for the SN progenitors on 
pre-SN images.

\acknowledgements

Financial support for this work was provided by NASA through Grant Nos.~AR-5793, AR-6371, AR-8006, and 
GO-6043 from the Space Telescope Science Institute, which is operated by AURA, Inc., under
NASA Contract No.~NAS 5-26555; we also acknowledge
NSF grant AST-9417213.  
We are grateful to Jay Anderson for assisting us with measuring the brightness of SN 1994D.
We appreciate fruitful consultations with Craig Sosin and Adrienne Cool.  We also appreciate assistance
from John Biretta at STScI on WFPC2 image header coordinate information.  SVD is grateful to 
the Division of Astronomy \& Astrophysics at UCLA, especially to Jean Turner.
This research has made use of the NASA/IPAC Extragalactic Database (NED) which is 
operated by the Jet Propulsion Laboratory, California Institute of Technology, under 
contract with the National Aeronautics and Space Administration.

\clearpage

\begin{deluxetable}{lcccclc}
\def\phmm{\phm{$-$}}
\tablenum{1}
\tablecolumns{7}
\tablewidth{7in}
\tablecaption{Summary of Archival Data}
\tablehead{\colhead{Galaxy} & \colhead{SN} & \colhead{Type} & \colhead{Detector} & 
\colhead{Filter} & \colhead{UT Date} & \colhead{Exposure} \nl
\colhead{} & \colhead{} & \colhead{} & \colhead{} & \colhead{} & \colhead{} & 
\colhead{(s)}}
\startdata
NGC 253  & 1940E & I & WF4 & F675W & 1994 Sep 17 & 400 \nl
                             &&&& F656N &             & 2400 \nl
NGC 891  & 1986J & IIn & WF2 & F606W & 1994 Dec 1 & 160 \nl
NGC 1003 & 1937D & Ia & WF4 & F606W & 1994 Nov 5 & 160 \nl
NGC 1275 & 1968A & I & WF2 & F606W & 1994 Mar 31 & 560 \nl
NGC 1365 & 1983V & Ic & WF3 & F547M & 1995 Jan 15 & 200 \nl
                              &&&& F555W &             & 300 \nl
                              &&&& F814W &             & 300 \nl
NGC 1614 & 1996D & Ib/c & PC & F606W & 1994 Dec 11 & 500 \nl
NGC 3227 & 1983U & Ia & PC & F606W & 1995 Feb 23 & 500 \nl  
NGC 3627 & 1989B & Ia & WF2 & F606W & 1994 Dec 28 & 160 \nl
         & 1997bs & IIn & WF4 &&& \nl
NGC 3690 & 1993G & II & WF2 & F606W & 1994 Sep 17 & 500 \nl
& 1998T & Ib & PC \nl
& 1999D & II & WF4 \nl
NGC 4258 & 1981K & II & WF3 & F547M & 1995 Mar 16 & 1160 \nl
                              &&&& F502N &             & 2300 \nl
                              &&&& F658N &             & 2300 \nl
NGC 4303 & 1926A, & II-L & WF4 & F606W & 1994 Jun 6 & 160 \nl
& 1964F & II & WF2 &&& \nl
NGC 4321 & 1959E & I & WF2 & F555W & 1994 Apr 23--Jun 19 & 21600 \nl
                                          &&&& F814W &                     & 7200 \nl
NGC 4490 & 1982F & II-P & PC & F606W & 1994 Dec 3 & 160 \nl
NGC 4496A/B & 1960F & Ia & WF4 & F555W & 1994 May 27--Aug 7 & 68000 \nl
                                                 &&&& F814W &         & 16000 \nl
& 1988M & II & WF2 &&& \nl
NGC 4526 & 1994D & Ia & PC & F555W & 1994 May 9 & 60, 230 \nl
                              &&&& F814W &            & 60, 230 \nl
NGC 4536 & 1981B & Ia & WF4 & F555W & 1994 May 12 & 660 \nl
                              &&&& F675W &             & 280 \nl
                              &&&& F814W &             & 660 \nl
                              &&&& F658N &             & 3000 \nl
NGC 4618 & 1985F & Ib & PC & F606W & 1995 Feb 16 & 160 \nl
NGC 4621 & 1939B & I & WF4 & F555W & 1995 Feb 5 & 1050 \nl
                             &&&& F814W &            & 1050 \nl
NGC 4651 & 1987K & IIb & WF2 & F555W & 1994 May 20 & 600 \nl
                               &&&& F814W &             & 600 \nl
                     &&& PC & F218W & 1995 Mar 4 & 1800 \nl
                               &&&& F547M &            & 300 \nl
NGC 4725 & 1940B & II-P & WF2 & F439W & 1995 Apr 12--Jun 15 & 2500 \nl
                                &&&& F555W &                     & 30000 \nl
                                &&&& F814W &                     & 10000 \nl
NGC 4772 & 1988E & Ipec & WF3 & F606W & 1994 Aug 13 & 160 \nl
NGC 5195 & 1945A & I & PC & F218W & 1994 Oct 8 & 1400 \nl
                             &&&& F547M &            &  230 \nl
NGC 5253 & 1895B & Ia & WF3 & F255W & 1995 May 29 & 6900 \nl
                    &&& PC & F606W & 1995 Mar 11 & 500 \nl
\tablebreak
NGC 5427 & 1976D & Ia & WF4 & F606W & 1994 Jun 25 & 500 \nl
NGC 5597 & 1981E & II & WF2 & F606W & 1994 Jul 15 & 500 \nl
NGC 6221 & 1990W & Ic & PC & F606W & 1995 Apr 2 & 500 \nl
NGC 7742 & 1993R & Ipec & WF3 & F336W & 1995 Jul 9 & 2100 \nl
                                &&&& F555W &            & 480 \nl
                                &&&& F675W &            & 480 \nl
                                &&&& F814W &            & 680 \nl
\enddata
\end{deluxetable}

\clearpage

\begin{deluxetable}{cccccc}
\def\phmm{\phm{$-$}}
\tablenum{2}
\tablecolumns{3}
\tablewidth{6in}
\tablecaption{Conversions from Johnson-Cousins to WFPC2 Synthetic Magnitudes and Colors\tablenotemark{a}}
\tablehead{\colhead{WMAG or WCOL} & \colhead{JMAG} & \colhead{JCOL} & \colhead{$C_1$} & 
\colhead{$C_2$} & \colhead{$Z$}} 
\startdata
F555W & $V$ & $B-V$ & 0.066 & $-$0.026 & $-$0.004 \nl
F336W$-$F439W & \nodata & $U-B$ & 1.204 & $-$0.233 & $-$0.074 \nl
F439W$-$F555W & \nodata & $B-V$ & 1.024 & $+$0.106 & $+$0.016 \nl
F555W$-$F675W & \nodata & $V-R$ & 1.329 & $-$0.152 & $-$0.003 \nl
F555W$-$F814W & \nodata & $V-I$ & 1.043 & $-$0.005 & $-$0.014 \nl
\enddata
\tablenotetext{a}{Based on either the expression WMAG = JMAG$\ +\ C_1$JCOL$\ + C_2$JCOL$^2\ +\ Z$ or
WCOL=$C_1$JCOL$\ +\ C_2$JCOL$^2\ +\ Z$ and derived via a uniformly weighted least-squares fit.}
\end{deluxetable}

\begin{deluxetable}{ccccc}
\def\phmm{\phm{$-$}}
\tablenum{3}
\tablecolumns{3}
\tablewidth{6in}
\tablecaption{Extinction and Reddening for Johnson-Cousins and WFPC2 Bandpasses\tablenotemark{a}}
\tablehead{\colhead{Band(s)} & \colhead{$C_1$} & \colhead{$C_2$}} 
\startdata
$A_U$ & 4.794($\pm$0.005) & $-$0.011($\pm$0.001) \nl
$A_B$ & 4.073($\pm$0.006) & $-$0.061($\pm$0.001) \nl
$A_V$ & 3.148($\pm$0.002) & $-$0.027($\pm$0.001) \nl
$A_R$ & 2.615($\pm$0.004) & $-$0.052($\pm$0.001) \nl
$A_I$ & 1.917($\pm$0.001) & $-$0.013($\pm$0.000) \nl
$A_{\rm F336W}$ & 5.591($\pm$0.053) & $-$0.454($\pm$0.011) \nl
$A_{\rm F439W}$ & 4.301($\pm$0.006) & $-$0.054($\pm$0.001) \nl
$A_{\rm F555W}$ & 3.199($\pm$0.004) & $-$0.046($\pm$0.001) \nl
$A_{\rm F675W}$ & 2.474($\pm$0.001) & $-$0.015($\pm$0.000) \nl
$A_{\rm F814W}$ & 1.910($\pm$0.003) & $-$0.033($\pm$0.001) \nl
\nl
$E(U-B)$ & 0.720($\pm$0.005) & $+$0.051($\pm$0.001) \nl
$E(V-R)$ & 0.533($\pm$0.001) & $+$0.025($\pm$0.000) \nl
$E(R-I)$ & 0.698($\pm$0.003) & $-$0.040($\pm$0.001) \nl
$E(V-I)$ & 1.231($\pm$0.002) & $-$0.015($\pm$0.000) \nl
$E({\rm F336W}-{\rm F439W})$\tablenotemark{b} & 1.055($\pm$0.026) & $-$0.165($\pm$0.013) \nl
$E({\rm F439W}-{\rm F555W})$ & 1.102($\pm$0.004) & $-$0.008($\pm$0.001) \nl
$E({\rm F555W}-{\rm F675W})$ & 0.725($\pm$0.003) & $-$0.032($\pm$0.001) \nl
$E({\rm F675W}-{\rm F814W})$ & 0.564($\pm$0.002) & $+$0.018($\pm$0.000) \nl
$E({\rm F555W}-{\rm F814W})$ & 1.288($\pm$0.002) & $-$0.013($\pm$0.001) \nl
\enddata
\tablenotetext{a}{Based on Cardelli et al.~(1989) reddening law, and of the form
$A_X$ or $E(X-Y)=C_1 E(B-V)+C_2 E(B-V)^2$.}
\tablenotetext{b}{This relation is highly nonlinear.  The coefficient values shown are 
appropriate only for stellar spectral types O through A with $0 \le E(B-V) \le 2$.}
\end{deluxetable}

\clearpage

\begin{figure*}
\figurenum{1}
%\plotone{vandyks.fig1.ps}
\caption{The environment of SN 1940B in NGC 4725 in a F555W WFPC2 image.
The SN position is within the 10\arcsec\ radius error circle.}
\end{figure*}

%\clearpage

\begin{figure*}
\figurenum{2}
\plottwo{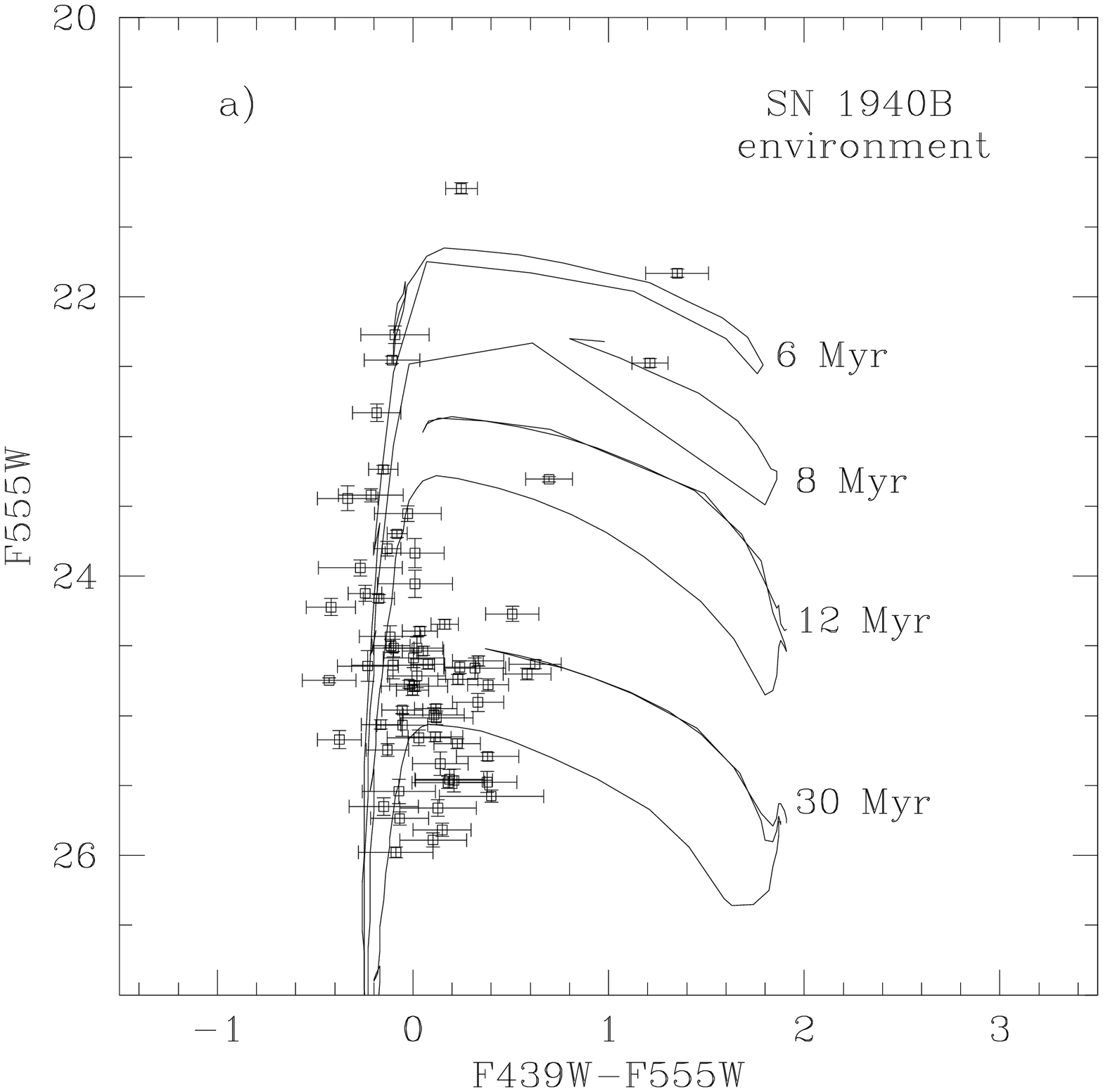}{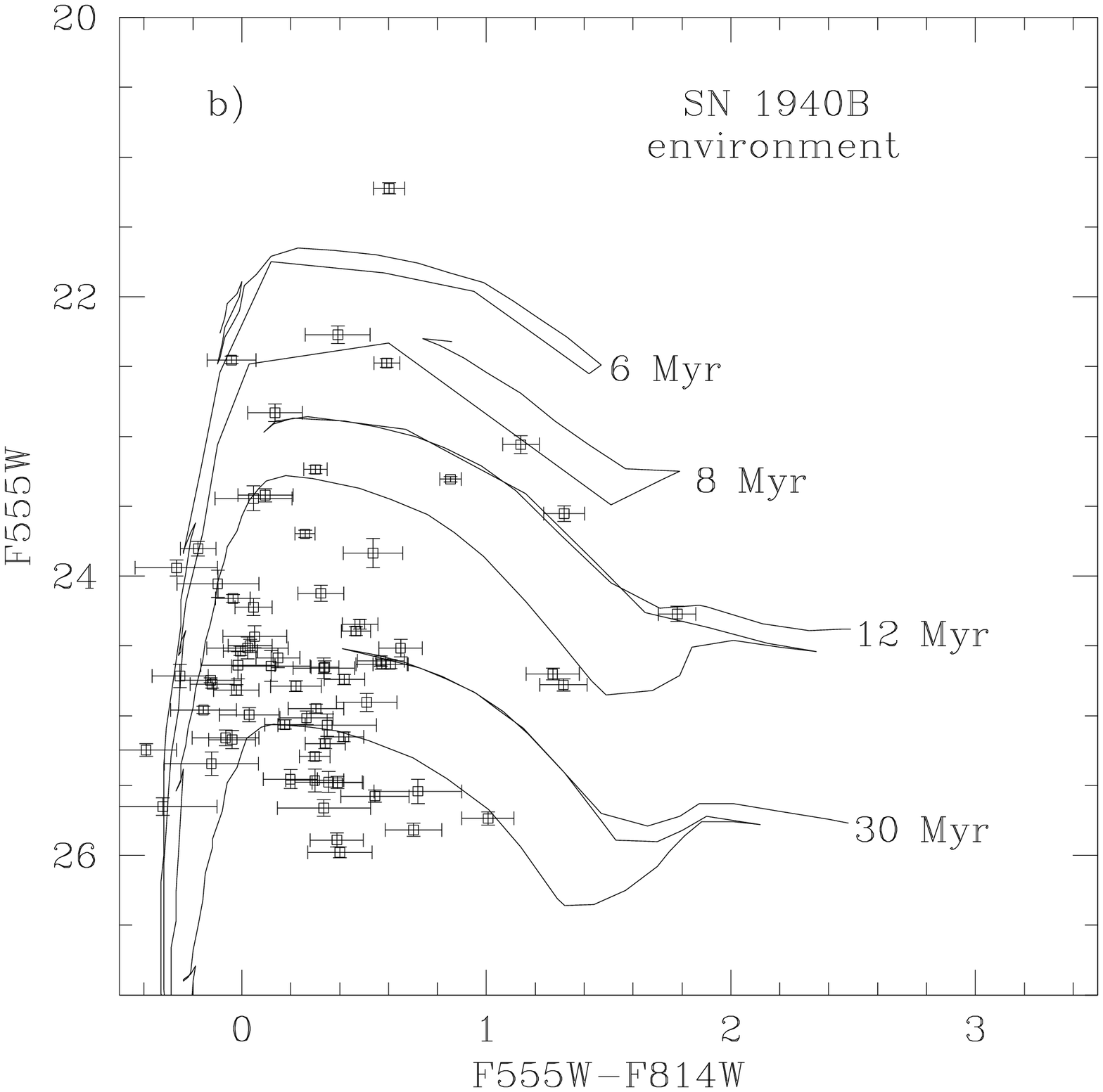}
\caption{The (a) F439W$-$F555W vs.~F555W and (b) F555W$-$F814W vs.~F555W
color-magnitude diagrams for the environment of SN 1940B in NGC 4725.  
Also shown are unreddened isochrones from B94, transformed to WFPC2 synthetic colors, 
and adjusted for the galaxy's distance modulus $m-M=30.50$ (Gibson et al.~1999).}
\end{figure*}

%\clearpage

%\begin{figure*}
%\figurenum{2}
%\plotone{vandyks.fig2b.ps}
%\caption{(Continued.)}
%\end{figure*}

%\clearpage

\begin{figure*}
\figurenum{3}
%\plotone{vandyks.fig3a.ps}
\caption{The environment of SN 1981K in NGC 4258, in {\it (a)} a F658N image, 
{\it (b)} a F502N image, and, {\it (c)} a F547M WFPC2 image.
The SN position is within the $1{\farcs}5$ radius error circle.}
\end{figure*}

\begin{figure*}
\figurenum{4}
%\plotone{vandyks.fig4.ps}
\caption{SN 1986J in NGC 891 and its environment in a F606W WFPC2 image.
The arrow points to the SN.}
\end{figure*}

%\clearpage

\begin{figure*}
\figurenum{5}
%\plotone{vandyks.fig5.ps}
\caption{The environment of SN 1987K in NGC 4651 in a F555W WFPC2
image.  The SN position is within the 2\arcsec\ radius error circle.}
\end{figure*}

%\clearpage

\begin{figure*}
\figurenum{6}
%\plotone{vandyks.fig6.ps}
\caption{The environment of SN 1988M in NGC 4496B in a F555W WFPC2 image.
The SN position is within the 5\arcsec\ radius error circle.}
\end{figure*}

%\clearpage

\begin{figure*}
\figurenum{7}
%\plotone{vandyks.fig7a.ps}
\caption{The environment of SN 1997bs in NGC 3627 (a) in a pre-explosion F606W 
WFPC2 image.  A $m_{\rm F606W}=22.86$ mag star is within the 1\arcsec\ radius 
error circle.  (b) The same field, but in a F555W image made as part of program
GO 6549 on 1997 Nov. 12.  The arrow points to the SN, which had 
$m_{\rm F555W}=21.42$ mag and is coincident with the star in (a).  (c) The
same as for (b), but in a F555W image made on 1998 Jan. 10, when the SN had 
faded to $m_{\rm F555W}=23.37$ mag.  The star seen in (a) is almost certainly the
SN progenitor.}
\end{figure*}

\begin{figure*}
\figurenum{8}
%\plotone{vandyks.fig8.ps}
\caption{The environment of SN 1983V in NGC 1365 in a F555W WFPC2 image.
The SN position is within the $1{\farcs}5$ radius error circle.}
\end{figure*}

%\clearpage

\begin{figure*}
\figurenum{9}
\plotone{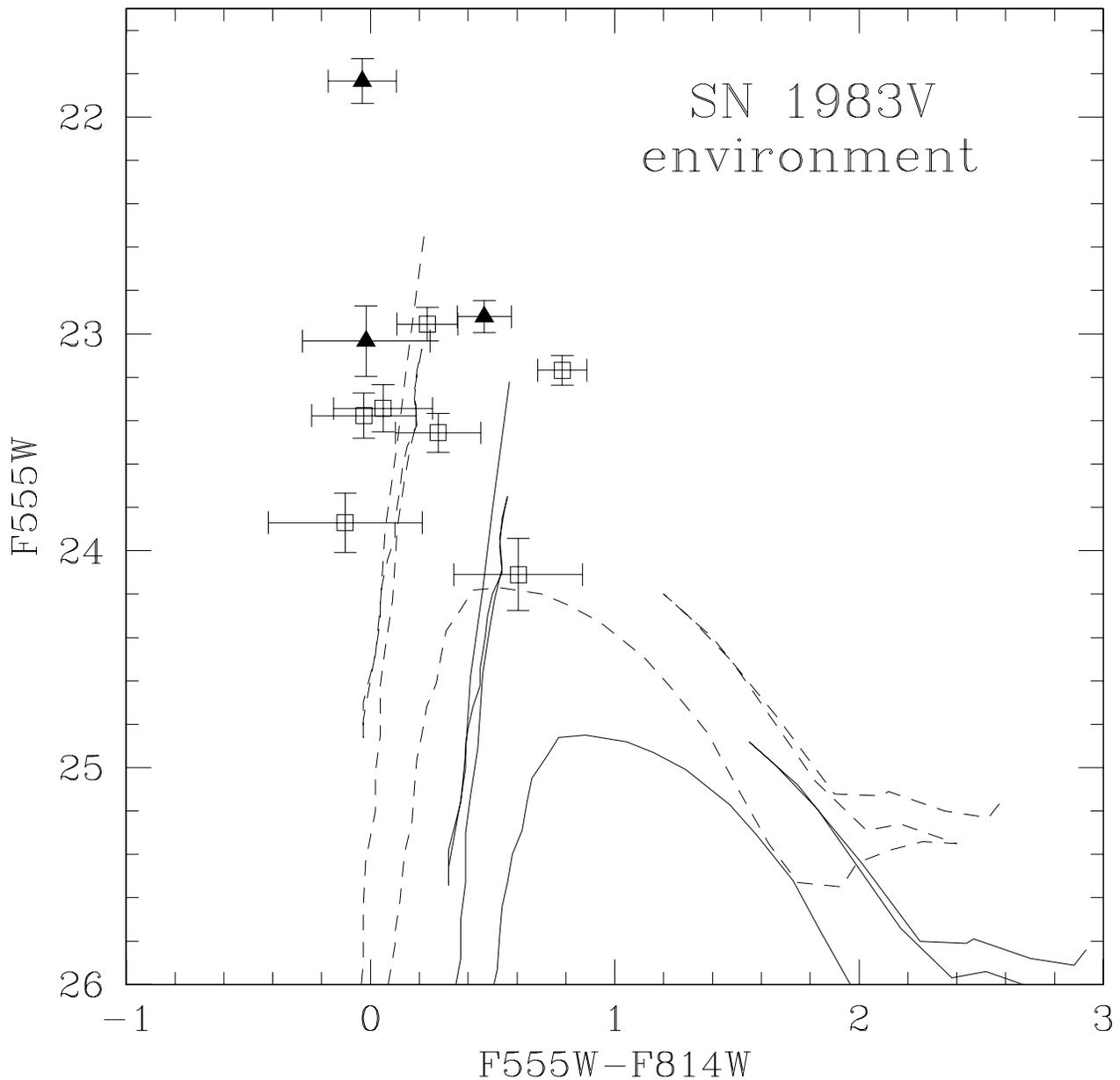}
\caption{The color-magnitude diagram for the environment of SN 1983V in NGC 1365.
Also shown are isochrones from B94 for 4 and 10 Myr, transformed to WFPC2 synthetic
colors, reddened by $E(B-V)\simeq 0.18$ ({\it dashed line}) and
$E(B-V)\simeq 0.4$ ({\it solid line}; see Clocchiatti et al.~1997), and adjusted for the 
galaxy's distance modulus $m-M=31.31$ (Silbermann et al.~1999).  
The {\it solid triangles\/} are those three objects
in or very near the error circle around the SN position.}
\end{figure*}

%\clearpage

\begin{figure*}
\figurenum{10}
\plotone{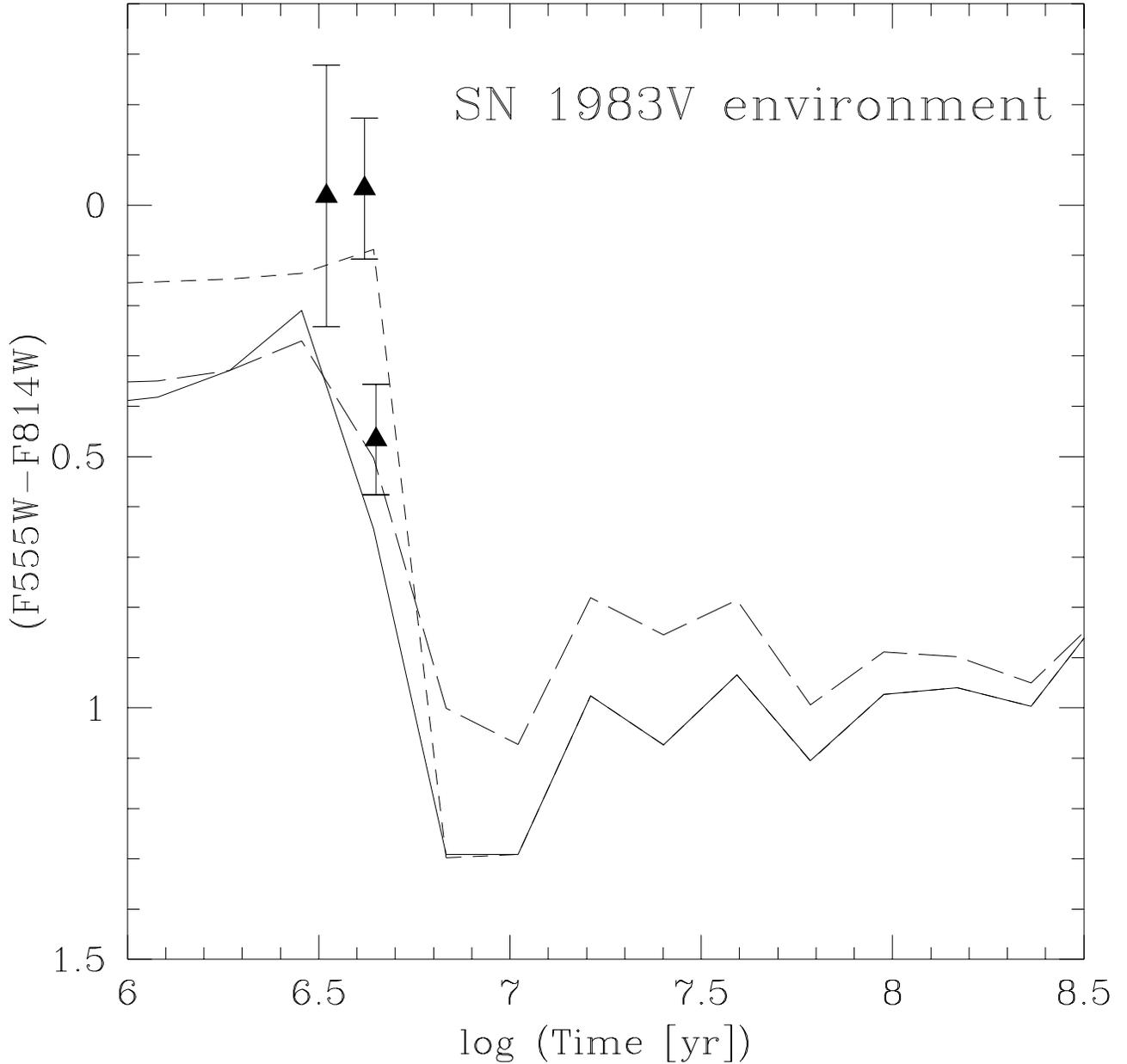}
\caption{The unresolved probable compact star clusters in the environment of SN 1983V in 
NGC 1365, compared to the color evolution models for an instantaneous starburst from
Leitherer \& Heckman (1995).  The line conventions are the same as their Figure 1;
the {\it solid line\/} is for a Salpeter ($\alpha=2.35$) initial mass function (IMF),
with a lower mass limit of 1 $M_{\sun}$ and an upper mass limit of 100 $M_{\sun}$; the
{\it long-dashed line\/} is for an IMF with a steeper slope ($\alpha=3.3$), with the 
same mass limits; the {\it short-dashed line\/} is for a Salpeter IMF with an upper mass 
limit of 30 $M_{\sun}$.  These models have been transformed to WFPC2 synthetic colors, and 
reddened by $E(B-V)\simeq 0.18$ (Clocchiatti et al.~1997).  The bright blue clusters are 
consistent with ages of $\lesssim$5 Myr.}
\end{figure*}

%\clearpage

\begin{figure*}
\figurenum{11}
%\plotone{vandyks.fig11.ps}
\caption{The environment of SN 1985F in NGC 4618 in a F606W WFPC2 image.
The SN position is within the $\sim$2\arcsec\ radius error circle.}
\end{figure*}

%\clearpage

\begin{figure*}
\figurenum{12}
%\plotone{vandyks.fig12.ps}
\caption{The environment of SN 1990W in NGC 6221 in a F606W WFPC2 image.
The SN position is within the $1{\farcs}5$ radius error circle.}
\end{figure*}

%\clearpage

\begin{figure*}
\figurenum{13}
%\plotone{vandyks.fig13.ps}
\caption{The environment of SN 1996D in NGC 1614 in a F606W WFPC2 image.
The SN position is within the 2\arcsec\ radius error circle.}
\end{figure*}

%\clearpage

\begin{figure*}
\figurenum{14}
%\plotone{vandyks.fig14.ps}
\caption{The environment of SN 1998T in NGC 3690 in a pre-explosion F606W WFPC2 image.
The SN position is within the 2\arcsec\ radius error circle.}
\end{figure*}

%\clearpage

\begin{figure*}
\figurenum{15}
%\plotone{vandyks.fig15.ps}
\caption{The environment of SN 1940E in NGC 253 in a F656N WFPC2 image.
The SN position is within the 10\arcsec\ radius error circle.}
\end{figure*}

%\clearpage

\begin{figure*}
\figurenum{16}
%\plotone{vandyks.fig16.ps}
\caption{The environment of SN 1959E in NGC 4321 (M100) in a F555W WFPC2 image.
The SN position is within the $1{\farcs}5$ radius error circle.}
\end{figure*}

%\clearpage

\begin{figure*}
\figurenum{17}
%\plotone{vandyks.fig17.ps}
\caption{The color-magnitude diagram for the larger environment of SN 1959E in NGC 4321 
(M100).  Also shown are isochrones from B94 for 6, 10, 20, 50, and 100 Myr, transformed to 
WFPC2 synthetic colors, reddened by $E(B-V)=0.01$, and adjusted by the galaxy's distance 
modulus $m-M=31.04$ (Ferrarese et al.~1996).  The {\it open triangle\/} indicates the 
magnitude and color of the background within the error circle shown in Figure 16.}
\end{figure*}

%\clearpage

\begin{figure*}
\figurenum{18}
%\plotone{vandyks.fig18.ps}
\caption{The environment of SN 1960F in NGC 4496A in a F555W WFPC2 image.
The SN position is within the $1{\farcs}5$ radius error circle.}
\end{figure*}

\clearpage

\begin{figure*}
\figurenum{19}
%\plotone{vandyks.fig19.ps}
\caption{The color-magnitude diagram for the larger environment of SN 1960F in NGC 
4496A.  Also shown are isochrones from B94 for 4, 10, 20, 50, 100, 300 Myr, and 
1 Gyr, transformed to WFPC2 synthetic colors, reddened by 
$E(B-V)=0.05$, and adjusted by the galaxy's distance modulus $m-M=31.03$
(Saha et al.~1996a).  The {\it open triangle\/} indicates the magnitude and color of the
background within the error circle shown in Figure 18.}
\end{figure*}

%\clearpage

\begin{figure*}
\figurenum{20}
%\plotone{vandyks.fig20a.ps}
\caption{The environment of SN 1981B in NGC 4536 in {\it (a)} a F555W image;
and, {\it (b)} a F658N WFPC2 image.
The SN position is within the $1{\farcs}5$ radius error circle.}
\end{figure*}

\begin{figure*}
\figurenum{21}
%\plotone{vandyks.fig21.ps}
\caption{The environment of SN 1989B in NGC 3627 in a F606W WFPC2 image.
The SN position is within the 2\arcsec\ radius error circle.}
\end{figure*}

%\clearpage

\begin{figure*}
\figurenum{22}
%\plotone{vandyks.fig22.ps}
\caption{The environment of SN 1993R in NGC 7742 in a F555W WFPC2 image.
The SN position of the SN is within the 1\arcsec\ radius error circle.}
\end{figure*}

%\clearpage

\begin{figure*}
\figurenum{23}
%\plotone{vandyks.fig23.ps}
\caption{The environment of SN 1994D in NGC 4526 in a F555W WFPC2 image.  
The SN is indicated by the arrow, just north of the galaxy's prominent dust lane.  
The horizontal bar in the lower left represents 5\arcsec.}
\end{figure*}

\end{document}